# Thermophysical properties of FLiBe using moment tensor potentials

*Siamak Attarian [a*], Dane Morgan [a*], Izabela Szlufarska [a*]*

[a] Department of Materials Science and Engineering, University of Wisconsin, 1509 University Ave, Madison, WI, 53706, USA

* sattarian@wisc.edu

* ddmorgan@wisc.edu

* szlufarska@wisc.edu

**Abstract:**

Fluoride salts are prospective materials for applications in some next-generation nuclear reactors and their thermophysical properties at various conditions are of interest. Experimental measurement of the properties of these salts is often difficult and, in some cases, unfeasible due to challenges from high temperatures, impurity control, and corrosivity. Therefore, accurate theoretical methods are needed for fluoride salt property prediction. In this work, we used moment tensor potentials (MTP) to approximate the potential energy surface of eutectic FLiBe (66.6% LiF – 33.3% $BeF_2$) predicted by the *ab initio* (DFT-D3) method. We then used the developed potential and molecular dynamics to obtain several thermophysical properties of FLiBe, including radial distribution functions, density, self-diffusion coefficients, thermal expansion, specific heat capacity, bulk modulus, viscosity, and thermal conductivity. Our results show that the MTP potential approximates the potential energy surface accurately and the overall approach yields very good agreement with experimental values. The converged fitting can be obtained with less than 600 configurations generated from DFT calculations, which data can be generated in just 1200 core hours on today's typical processors. The MTP potential is faster than many machine learning potentials and about one order of magnitude slower than widely used empirical molten salt potentials such as Tosi/Fumi.

## 1. Introduction:

Molten salts have garnered significant attention due to their applications in molten salt reactor systems both as coolants and fuel salts [1], concentrated solar power plants [2], and molten salt batteries [3]. In particular, FLiBe has been used in the molten salt reactor experiment in the 1960s [4] and it is one of the prospective salts to be used in generation IV reactors [5]. FLiBe has a low neutron absorption cross-section, high volumetric heat capacity [6], and is liquid in the temperature range 732 K - 1703 K [7]. Although many experimental measurements of FLiBe are available in the literature, the reported thermophysical properties are scattered with uncertainties up to 20% [8]. This scatter in measured properties has been suggested to be due to such issues as the presence of impurities and deviation from the 2:1 ratio between LiF and $BeF_2$ in the experiment [8]. As an alternative, accurate theoretical methods can be utilized to calculate the thermophysical properties of FLiBe in well-controlled conditions.

Several works based on *ab initio* molecular dynamics (AIMD) have been conducted to study various aspects of FLiBe [9–14]. While AIMD simulation is a highly accurate method based on quantum mechanics, a number of properties relevant for applications, such as viscosity, thermal conductivity, melting temperature, etc. are impractical to study with AIMD due to the limitations in length- and time-scales of such simulations. An alternative approach involves using interatomic potential molecular dynamics (IPMD) simulations [15–18] where the interactions between atoms are fit to experimental and quantum mechanical calculations. IPMD, also often called classical MD, allows simulations of larger supercells for longer times than AIMD, which times and sizes are sufficient for predicting complex thermodynamic and transport properties. Properties from classical MD simulations can generally be determined with enough numerical precision that their errors are dominated by the accuracy of the underlying interatomic potentials.

In the past decade, significant progress has been made in the development of so-called machine learning interatomic potentials (MLIP), where the functional form of the potential does not have a physical meaning as in the case of traditional potentials and the potential parameters are fitted using ML-based techniques. MLIPs have shown promise for conducting MD simulations with near *ab initio* accuracy but on time- and length-scales comparable to traditional interatomic potentials. Various forms of MLIPs are currently in use in the materials science community, including neural networks interatomic potentials (NNIP) [19], Gaussian approximation potentials (GAP) [20], deep potentials (DPMD) [21], spectral neighbor analysis potentials (SNAP) [22], moment tensor potentials (MTP) [23], etc. MLIPs have also found applications in molten salt modeling. For example, Liang *et al.* [24] used DPMD to model $MgCl_2$-KCl and found that the $Mg^{2+}$ ions in this system have a distorted tetrahedral local geometry. Their reported thermal expansion coefficient and viscosity were in good agreement with the experiments. Sivaraman *et al.* [25] used the GAP potential to model molten LiCl. Their calculated density and self-diffusion coefficients also agreed well with experiments. In another example, Feng *et al.* [26] used DPMD [27] to study the structure of molten $LaCl_3$. They found that molten $LaCl_3$ mainly consists of sevenfold and eightfold coordinated structures. In a recent article, Rodriguez *et al.* [28] used the DPMD framework to develop a potential for LiF and FLiBe. Their results were in good agreement with AIMD simulations, however, as the authors themselves noted, due to the lack of van der Waals dispersion interactions in their *ab initio* simulations, the simulation results deviated from experimental values. Most of the published works on MLIPs for molten salts have used NNIP or DPMD. The number of data points that are typically used to train such potentials is in the tens of thousands [26,29–32] and in some cases hundreds of thousands [28,33–35]. Here, one data point consists of a single energy and a set of force and stress vectors on atoms from one periodic unit cell configuration (the unit cell typically contains about 100 atoms). Recent work [36] has shown that MTP can be trained with much smaller datasets and is much faster compared to other MLIPs for MD simulations. Up until now, the applicability of MTP potentials to molten salts has not been demonstrated.

Here, we use MTP to approximate the potential energy surface of FLiBe, predicted by the DFT-D3 [37] method, which considers dispersion corrections. We then use the developed potential to simulate FLiBe and to predict its properties. There are two main results of this paper: 1) We assess the ability of MTP potentials to model FLiBe, which is essentially an ionic compound, and we consider such factors as the accuracy of fitting, training data requirements, and resulting MLIP speed; 2) We assess the ability of DFT with dispersion corrections combined with MLIP to accurately predict thermophysical properties of FLiBe by comparing our model results with available experimental values.

## 2. Methodology:

### 2.1. MTP potential

The total energy of a system in MTP potential is calculated by the sum of the energies of individual atoms:

$$E = \sum_{i=1}^{N_{tot}} E_i \qquad Eq.1$$

$E_i$ is the energy of the $i^{th}$ atom, which is calculated as

$$E_i = \sum_{\alpha} \xi_\alpha B_\alpha(n_i) \qquad Eq.2$$

Here, $n_i$ is determined based on the atomic environment around atom $i$, $B_\alpha$'s are basis functions that are defined by contraction of moment tensors, $\xi_\alpha$'s are fitting parameters, and $\alpha$ is the number of basis functions. The moment tensors are defined as

$$M_{\mu,\nu}(n_i) = \sum_{j} f_\mu(|r_{ij}|, z_i, z_j) \boldsymbol{r_{ij}} \otimes ... \otimes \boldsymbol{r_{ij}} \qquad Eq.3$$

In the above equation, the summation is over $j$ neighboring atoms that fall within the cutoff radius centered at atom $i$. $z_i$ and $z_j$ are the types (atomic species) of atoms $i$ and $j$, $r_{ij}$ is the relative position of atom $j$ with respect to atom $i$ ($r_{ij}=r_j-r_i$), and the outer product of vector $r_{ij}$ is done $\nu$ times. The $f_\mu$ function is the radial part of the moment tensor and it is calculated based on a series of fitting parameters $c_\mu$ and Chebyshev polynomials. Interested readers are referred to Refs. [23,38] for the complete description of MTP.

In this work, we used the MLIP package [38] to fit the parameters of MTP and we used its library developed for the LAMMPS package [39] to perform MD simulations. The MLIP package contains a series of MTP potentials with preset hyperparameters named MTP level 2 (MTP02), level 4 (MTP04), etc., that we tested for potential fitting. By increasing each MTP level, the complexity of the potential increases, which means that more fitting parameters are introduced to the potential. We assessed the performance of the MTP potential in three stages. In the first stage, we used a large training set to compare the computational cost vs accuracy of each MTP level. After comparing between different MTP levels we chose the most efficient MTP level and used that for the subsequent stages. In the second stage, we tested the effect of the training set size on the predicted energies and forces. In the final stage, we used an active learning scheme based on the concept of D-optimality [38] implemented in the MLIP package to create a smaller training data set that would provide the desired accuracy.

To combine the errors in energies, forces, and stresses during the fitting procedure, the fitting weights of 1, 0.01, and 0.001 were used, respectively, which are the default weights in the MLIP code and have shown to work well in previous studies [38,40].

## 2.2. Data generation and fitting procedure

All the atomic configurations used in this work as training and testing data were obtained from AIMD simulations or single-point energy calculations of supercells containing 98 atoms (28 Li, 14 Be, and 56 F). DFT calculations were performed using the VASP 5.4.4 package [41] in the canonical (NVT) ensemble using the Nosé thermostat [42] and by considering spin polarization. PBE-GGA approximation [43] was used for the exchange-correlation functional and the D3 method of Grimme [37] was used to account for dispersion forces. PAW-PBE potentials which were used in this study are Li_sv ($1s^2 2s^1$), Be ($2s^2$), and F ($2s^2 2p^5$). An energy cutoff of 600 eV was used for the plane-wave basis set and a single gamma point was used to sample the Brillouin zone.

In the first stage of our assessment of the MTP potential, we compared the different MTP potential levels. Initially, the Packmol package [44] was used to generate a random atomic configuration based on the experimental density (2.01 g/cm$^3$ at 823 K [45]). To save time in reaching the equilibrium structure of FLiBe, we used this configuration to initialize classical MD simulations with an existing potential in the NVT ensemble at temperatures 823 K, 973 K, 1223 K, and 1423 K, for 10 ps at each of these temperatures. At each temperature, the system was found to reach equilibrium after about 2 ps of the simulation as after 2 ps the pressure and the temperature of the system showed minor oscillations around the average values. The final configurations from each of the above NVT MD simulations were used as the starting points for AIMD simulations. The interatomic potential parameters for FLiBe were obtained from [15] by assuming a Tang–Toennies dispersion damping of 1.0 and using the parameters in the Tosi/Fumi potential. This step of using a classical MD potential to generate input for AIMD simulations is not strictly necessary for fitting the ML potential and it is used mainly to accelerate the equilibration of atomic configuration used in AIMD. Several AIMD simulations were performed at four temperatures: 823 K, 973 K, 1223 K, and 1423 K, and at different densities. At the equilibrium density for each temperature, we ran an AIMD simulation in the NVT ensemble for 1.5 ps with a timestep of 1 fs. Separate AIMD simulations in the NVT ensemble at densities ±3% and ±6% higher/lower than the experimental density were also performed at each temperature for 0.5 ps to add more diversity to the training data. We collected the atomic configurations at each time step and overall, 14,000 atomic configurations were obtained. Out of these, we picked the atomic configurations at every 10$^{th}$ timestep to include in the training set (1,400 configurations) and the remaining configurations were used as an initial testing set called Test1 (12,600 configurations). We used the training set to fit MTP potentials with different levels and compared the accuracies and computation times of the fitted potentials.

We also generated a more demanding test data set with less correlation to the training data, which we call Test2. To generate Test2 we obtained a random atomic configuration from Packmol and conducted an ionic relaxation using VASP. Then we started from the relaxed system and ran two AIMD simulations one at 600 K at a density 10% lower than the experimental density and one at 1600 K at a density 10% higher than the experimental density, each for 5 ps. Due to the different temperatures and densities of these simulations compared to the simulations used to generate the training data, these runs were more likely to have atomic configurations much different than what was used in the training data, which makes this a more demanding data set for assessing the potential. Overall, 10,000 configurations were generated for this Test2 data set.

In the second stage of our assessment, we examined the effect of the training set size (the number of training data in the training set) on the accuracy of the MTP potential. Toward this end, we made many random subsets of the original 1,400 training set with various set sizes, fit the potential with these subsets, and assessed them with the Test1 and Test2 sets.

In the final stage of our assessment, we developed a potential based on the active learning scheme. To do that, we conducted a new AIMD simulation and a new set of single-point energy calculations based on the actively selected configurations from MD simulations using the D-optimality criterion [46]. The active learning method is discussed in more detail in section 3.3. This final potential was then assessed with the Test1 and Test2 sets.

Since FLiBe is an ionic compound long-range interactions are important in calculating the energies. We have tested the convergence of the energies and forces with respect to the cutoff radius of the interactions and found that there are no improvements in the energy and force errors beyond the cutoff radius of 7 Å. However, there are still some structural features detectable in the radial distribution functions of the salt (shown in Figure A.1 in the supplementary data) up to the cutoff of 10 Å. We have therefore used 10 Å as a cutoff in our final energy calculations. We note that this conservative choice was practical in this work, but others might want to consider a smaller cutoff as the MD simulations with the cutoff of 7 Å are 2-3 times faster than for 10 Å. This distance is long enough to include multiple neighbor shells in the liquid, and we expect that electrostatic interactions are negligible beyond this distance for near-equilibrium configurations of FLiBe. Subtracting off the long-range electrostatic interactions in advance to assure their convergence could potentially reduce the range needed or increase accuracy, but we have not pursued this strategy here.

**2.3. Molecular dynamics**

All the MD simulations in this work were performed using the LAMMPS package [39]. The initial atomic positions of each MD simulation were generated randomly using Packmol. We needed to make sure that when we start to calculate a property of FLiBe, the system is in equilibrium. To this end, we started each MD simulation by assigning random velocities from the Maxwell - Boltzmann distribution at 1600 K and let the system cool down for 10 ps to the temperature and the pressure relevant to that simulation. For example, if we needed to calculate a property of FLiBe at 800 K and 1 kbar, first we performed a controlled pressure-controlled temperature (NPT) simulation that took the system from 1600 K and 0 bar to 800 K and 1 kbar during 10 ps, and then continued the simulation at 800 K and 1 kbar to calculate the desired property. For the remainder of the paper, we will refer to this stage of each MD simulation as the initial equilibration.

In order to calculate radial distribution function (RDF), volume, density, diffusivity, enthalpy, thermal expansion coefficient, and specific heat capacity, we used a simulation cell consisting of 6272 atoms and a simulation time step of 1 fs. After the initial equilibration, we kept the system in the NPT ensemble at $T_d$ and 0 bar for 100 ps where $T_d$ is the desired temperature in each simulation. The quantity of interest was obtained by averaging over the full production run time of 100 ps. The radial distribution function (RDF) was calculated at just 973 K. For the other properties we performed simulations at temperatures that ranged from 600 K to 1,600 K with 50 K increments, which resulted in a total of 21 simulations.

The self-diffusion coefficient was calculated from the slope of the mean squared displacement (MSD) using Einstein's relation [47]

$$D = \frac{1}{6} \lim_{t \to \infty} \frac{d}{dt} \left[ \frac{1}{N} \frac{1}{n_t} \sum_{i=1}^{N} \sum_{j=1}^{n_t} \left( r_i(t_j + dt) - r_i(t_j) \right)^2 \right] \qquad Eq.4$$

where $D$ is the self-diffusion coefficient, $N$ is the number of atoms, and $n_t$ is the number of time origins. The thermal expansion coefficient was calculated using the following equation

$$\alpha = \frac{1}{V}\left(\frac{dV}{dT}\right)\big|_p = -\frac{1}{\rho}\left(\frac{d\rho}{dT}\right)\big|_p \qquad Eq.5$$

where $V$ and $\rho$ are the equilibrium volume and density at each temperature, respectively. The specific heat was calculated using the following equation

$$c_p = \frac{\partial h}{\partial T} \qquad Eq.6$$

where *h* is the enthalpy.

The bulk modulus was calculated using a simulation setup the same as for the RDF and other properties described above, except for different temperatures and the use of a grid of pressures. Bulk modulus was calculated for 6 temperatures between 800 K and 1300 K at every 100 K. For each temperature ($T_d$), 7 different simulations were performed at pressures ($P_d$) ranging from -0.6 to 0.6 GPa chosen with 0.2 GPa intervals. After the initial equilibration, we kept the system in the NPT ensemble at constant temperature $T_d$ and constant pressure $P_d$ for 100 ps. In each of these simulations, the volume and the pressure were calculated as the average value over 100 ps. For each temperature, the bulk modulus was obtained by fitting the 3$^{rd}$ order Birch-Murnaghan equation of state [48,49] using volumes and pressures calculated from the 7 simulations performed with different $P_d$.

Viscosity was calculated using the Green-Kubo relation [50,51]

$$\eta = \frac{V}{k_B T} \int_0^\infty <P_{\alpha\beta}(t).P_{\alpha\beta}(0)> dt \qquad Eq.7$$

Here, $\eta$ is the viscosity, $k_B$ is the Boltzmann constant, and $P_{\alpha\beta}$ are the off-diagonal elements of the stress tensor. For viscosity calculations, we used a simulation cell consisting of 1,540 atoms and the simulation time step was 1 fs. An autocorrelation time of 20 ps was chosen for the viscosity calculations at 700 K and 750 K and 10 ps in the temperature range 800 K – 1200 K. In our simulations, the selected time intervals were shown to be sufficient to allow the decay of the autocorrelation function of the diagonal stress components to zero. At each temperature, after the initial equilibration, the simulation was carried out for 5 ns using the micro-canonical (NVE) ensemble. In all the simulations, the running integral of the stress autocorrelation function remained stable after about 4 ns.

Previous simulation studies on the thermal conductivity of molten salts have shown that it is a challenging quantity to calculate [15,52,53]. Here, we determined the thermal conductivity using the Muller-Plathe nonequilibrium method [54]. We follow the approach outlined by Pan *et al.* [55], where thermal conductivity calculated using the Muller-Plathe method was found to be in better agreement with experimental values than results based on the Green-Kubo method. In our calculations, we used a time step of 0.5 fs, a supercell containing 31,360 atoms (with dimensions 4.2 nm × 4.2 nm × 21 nm), and a kinetic energy swap rate of 1 in every 1,000 steps. For each temperature, after the initial equilibration in the NPT ensemble for 10 ps, the simulations ran for 2 ns in the NVE ensemble.

## 3. Results and discussion:

### 3.1. MTP potentials with increasing levels of complexity

Due to the nonlinearity of the MTP formulation for multicomponent systems, the fitting procedure is carried out using the Broyden-Fletcher-Goldfarb-Shanno (BFGS) method. The optimized parameters of the MTP potential obtained by this method depend on how the parameters have been initialized. This dependency implies that for the same training data, every optimization session can result in different optimized parameters, and therefore, different errors for forces and energies. Here, for each potential level, we carried out 3 optimization sessions and chose the one with the lowest combined errors (the potential levels and error weights are discussed in Section 2.1). Table 1 shows the root mean square errors (RMSE) of the energies and forces for the trained MTP potentials with different levels of complexity.

*Table 1. Root mean squared errors (RMSE) of energies and forces for different levels of the MTP potential. The numbers after MTP are the complexity levels of the potential as described in Section 2.1. The values are in the units of meV/atom for energies, and meV/Å for forces. The IPMD is based on the MD simulations with the Tosi/Fumi potential and is provided to compare the computational cost. Computational cost is shown in the units of (core seconds)/(atom.timestep), where atom.timestep is determined by dividing the total core seconds by the product of the number of atoms and the number of timesteps. Data is for an Intel Xeon Gold 5218R processor.*

|  | MTP06 | MTP08 | MTP10 | MTP12 | MTP14 | IPMD |
|---|---|---|---|---|---|---|
| Total number of fitting parameters | 149 | 153 | 232 | 245 | 340 | - |
| Energy RMSE (meV/atom) |  |  |  |  |  |  |
| Training set | 5.4 | 3.6 | 1.8 | 1.8 | 1.3 | - |
| Testing set (Test1) | 5.4 | 3.6 | 1.7 | 1.8 | 1.3 | - |
| Testing set (Test2 at 600 K) | 3.3 | 3.2 | 2.9 | 2.8 | 2.3 | - |
| Testing set (Test2 at 1600 K) | 5.4 | 3.8 | 2.5 | 2.1 | 2.5 | - |
| Force RMSE (meV/Å) |  |  |  |  |  |  |
| Training set | 173 | 111 | 59 | 59 | 41 | - |
| Testing set (Test1) | 173 | 111 | 59 | 59 | 41 | - |
| Testing set (Test2 at 600 K) | 159 | 95 | 53 | 58 | 39 | - |
| Testing set (Test2 at 1600 K) | 193 | 128 | 69 | 77 | 50 | - |
| Computational cost (core seconds)/(atom.timestep) | $8.7 \times 10^{-5}$ | $1.2 \times 10^{-4}$ | $2.2 \times 10^{-4}$ | $3.6 \times 10^{-4}$ | $6 \times 10^{-4}$ | $1.9 \times 10^{-5}$ |

The first noticeable result is that the errors of training and Test1 testing sets are almost the same for all the potentials. This suggests an excellent interpolating power of the MTP potentials which is the main expectation of the machine learning potentials. However, due to the correlation of the training and Test1 testing sets, one cannot be sure about the performance of the potential in relevant but unseen atomic environments. For this reason, we conducted 2 separate AIMD simulations at temperatures and densities different from the AIMD simulations that generated the training data to create the Test2 set as discussed in section 2.2. Between the two Test2 sets, the one at 1,600 K (Test2_1600) has higher errors compared to the errors of the training set, likely due to the faster ionic motion leading to the formation of more varied atomic environments within the supercells. Earlier studies [26,53,56] have shown that energy errors smaller than 5 meV/atom and force errors smaller than 100 meV/Å are generally sufficient for predicting such properties as density, RDF, diffusion coefficient, and viscosity. As can be seen in Table 1 MTP10, MTP12, and MTP14 provide sufficient accuracies by this measure for all the training and testing sets. Figure 1 shows the errors vs MTP level for the training set and Test2_1600 testing set.

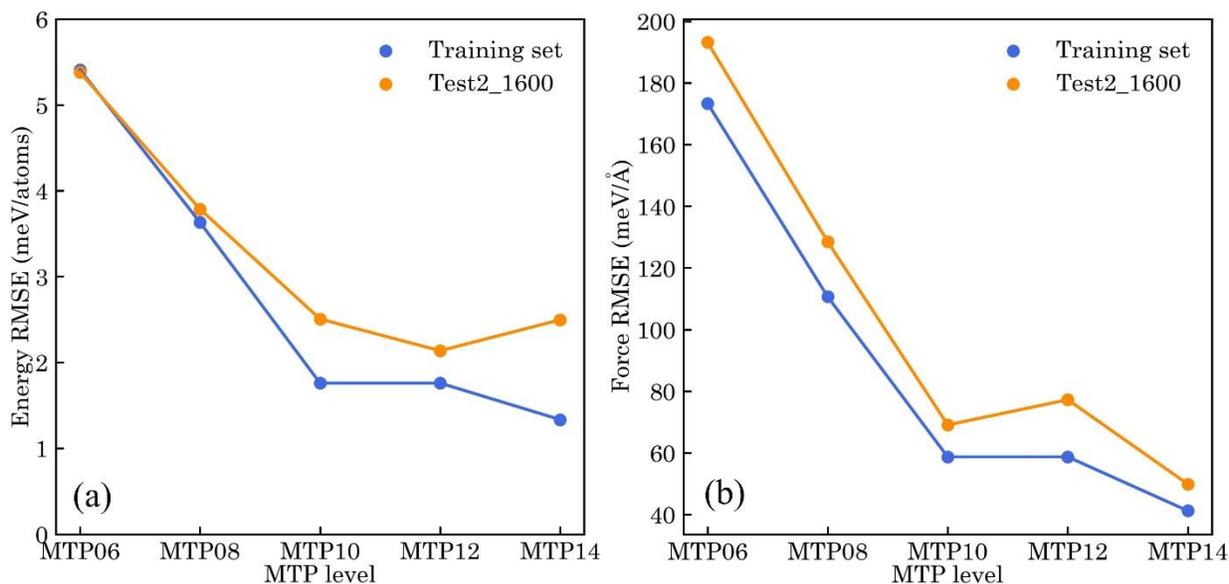

*Figure 1. Comparison of the root mean square errors (RMSE) for (a) energies and (b) forces of the training and testing sets predicted by different MTP levels.*

To better understand the role of the MTP potential complexity we consider its impact on the energy and force errors in Figure 1. Figure 1 (a) shows that from MTP06 to MTP10 the energy errors decrease considerably for both the training set and Test2_1600 set. At MTP12 the error does not change for the training set, but there is a slight decrease in the error of the Test2_1600 set. At MTP14 the error of the training set slightly decreases while the error of the Test2_1600 set increases. Figure 1 (b) shows that from MTP06 to MTP10 the force errors decrease considerably for both the training and Test2_1600 set. At MTP12 the error does not change for the training set, but the error of the Test2_1600 set increases. At MTP14 the errors of both curves slightly decrease. Figure 2 compares the energy errors of the training set vs computational cost for different MTP levels.

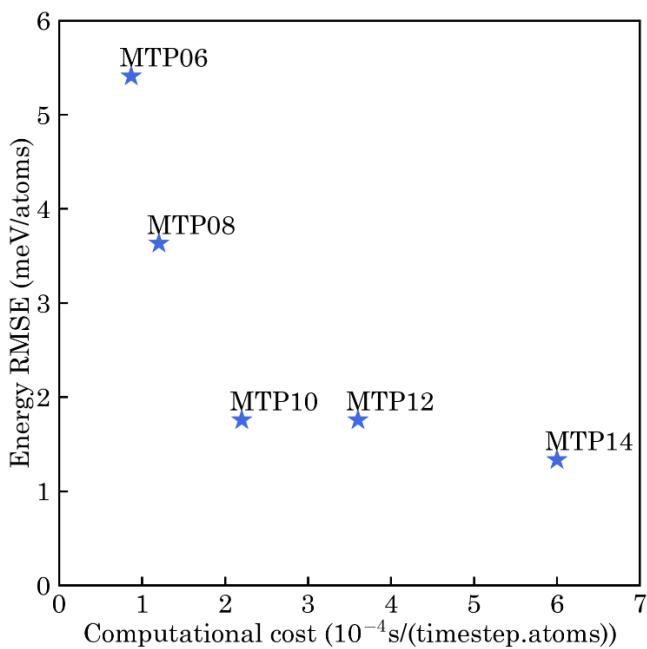

*Figure 2. Root mean square errors (RMSE) of the energies of the training set vs computational cost. The computational cost is reported for 1 core of the Intel Xeon Gold 5218R processor.*

Among the different levels of MTP potentials that showed acceptable errors, MTP10 was found to be the most efficient for our system of 3 species (Li, Be, F). MTP14 has slightly smaller errors, but it is almost 3 times slower than MTP10. As a result, we chose MTP10 for assessing the training set size effect and active learning in Sections 3.2 and 3.3.

It is interesting to consider the computational cost of MD simulations with MTP for FLiBe, which is in the range of $10^{-4}$ - $10^{-3}$ core seconds/(atom.timestep). This cost is lower than the NNIP potential developed for FLiBe in Ref. [31] which is in the range of $10^{-3}$ - $10^{-2}$ core seconds/(atom.timestep) or for the GAP potential fitted for molten $HfO_2$ [57] which is in the range $10^{-3}$ - $10^{-2}$ core seconds/(atom.timestep). We note that the potential development that was reported in Refs. [31] and [57] did not involve hyperparameter optimization. If such optimization was included (as in our work), the computational cost reported in those studies could potentially be reduced. However, in a study that was focused on the comparison between MLIPs [36], MTP was shown to be the fastest among the tested MLIP formulations (including NNIP and GAP). Nonetheless, our developed MTP potential is about an order of magnitude slower than the traditional IPMD with the Tosi/Fumi potential, as shown in Table 1.

**3.2. Training size effect**

The results of Section 3.1 showed that MTP potentials trained by a training set containing 10% of the configurations obtained from AIMD simulations can predict the energies and forces of the other 90% (testing set Test1) with the same error as the errors of the predictions of the training set. It is interesting to know the minimum training set size that would yield the same results. To that end, we fitted MTP10 potentials with various training set sizes. For each set size, we made 5 random samplings from the original training set (that contains 1400 atomic configurations), fitted a potential, calculated the errors, and averaged the results. Figure 3 shows the Learning curves for MTP10s with growing training set sizes.

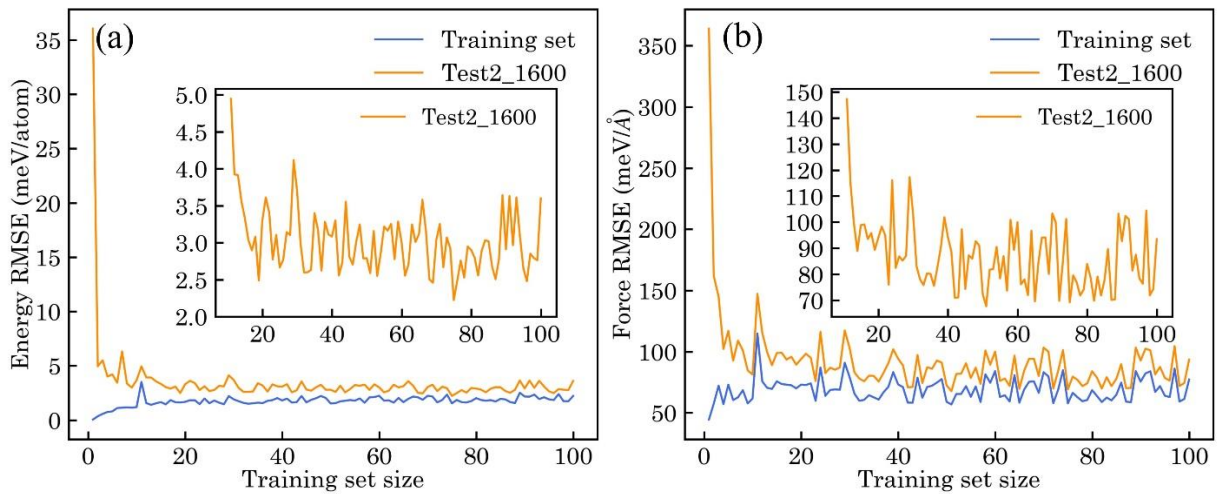

*Figure 3. Learning curves of MTP10 with growing training set sizes for (a) energies and (b) forces. The energy error of the test set stops decreasing and oscillates about 3 meV/atom after the training set size reaches the value of 20. The energy errors of the Test2_1600 set shown in the inset of (a) are magnified for the training set sizes between 10 and 100. In the case of the force errors (b), the decrease ceases around the training set size of 40 and starts oscillating about 85 meV/Å, as shown in the inset of (b).*

The insets in Figure 3 show the same data for the Test2_1600 set as in the main plot but starting from a training set size of 10. The fluctuations in the errors may partly be due to the uncertainties that arise from the fitting procedure (i.e., some fits may trap the parameters in a local optimum with a relatively higher error) and partly due to the quality of randomly sampled training data (i.e., the number of atomic environments that are represented by the data). The curves in the insets show that after about 40 training data the lowest values of the energy and force errors are relatively close to 2.5 meV/atom and 69 meV/Å, respectively, which are the errors of the MTP10 potential trained by the entire 1,400 training data. This trend suggests that it is possible to find a training set with as low as 40 samples and fit a potential with low prediction errors.

To test the stability of the MTP10 potentials fitted in this section we chose several of them that were fitted with training set sizes between 40 to 100 and had small testing errors. Using each of the selected potentials we conducted two MD simulations in the NPT ensemble one at 600 K and 0 bar and one at 1,600 K and 0 bar with a supercell containing 6,272 atoms. The MD simulations were run for 100 ps with a time step of 1 fs. Most of these MD simulations failed either due to the lost atoms error in LAMMPS or due to an extreme supercell expansion during the simulation. Only a few of the tested potentials successfully finished the MD simulations. It should be noted that the MPT10 potential that was fitted with the entire training set (1,400 training data) in Section 3.1 also passed this test. Based on these results we were not able to conclusively suggest a minimum number of randomly selected training data that would fit a stable MTP potential. Furthermore, this result suggests that low energy and force errors, even for a relatively independent test set like the Test2 set used here, do not imply that one has a stable potential that can be used robustly for practical molecular dynamics simulations even under the same temperature conditions as the test data.

## 3.3. Active selection of training samples

In the previous section, we showed that we can get a small testing error from a potential fitted with a few randomly sampled training data, however, such fitting does not necessarily result in a stable potential. An alternative method of selecting training data is through the process of active learning (active sampling) [58]. Active learning generally means selecting a set of training data that have enough diversity in the feature space for the purposes of fitting the potential. Several active learning schemes are in use for machine learning potential development [57,59,60]. Here, we used an active sampling method based on the D-optimality criterion as implemented in the MLIP package. In this method, a set of *n* configurations is initially generated and provided to the code as the starting point. Each configuration is converted to a set of *m* features and placed in a matrix form. The algorithm then selects *m* configurations out of *n* that have the maximum modulus of the determinant (|det(A)|). *m* is the number of the basis functions used in the MTP potential. These *m* configurations are called the active set. Later, when a new configuration is introduced to the algorithm, it is added to the training set, if it could increase |det(A)| by replacing one of the configurations already existing in the active set. [46]. This means that during the active learning procedure, we provide the algorithm with a set of configurations. Some of these configurations will be chosen to be added to the training set and the rest are discarded. Every time the training set is updated, the active set is also updated accordingly.

To fit a potential by active learning we generated an initial training set by running an AIMD simulation in the NVT ensemble at 1,100 K for 1 ps with a timestep of 1 fs and selecting 10% of the generated configuration (every 10th timestep). The details of the DFT calculations in this section (the number of atoms, energy cutoff, etc.) are the same as in Section 2.2. This data set was used both to fit an initial MTP10 potential and as the starting point of the active learning procedure. Using the initial potential, we then ran three MD simulations with a small supercell (98 atoms) in the NPT ensemble at temperatures 1,000 K, 1,100 K, and 1,200 K, each for 20 ps with a time step of 1 fs, and collected 10% of the configurations (every 10th timestep for each simulation). The active learning algorithm selected 315 out of the 6,000 collected configurations to be added to the training set using the procedure discussed in the preceding paragraph. We conducted single-point DFT calculations with the selected configurations and added the results to the training set. We then fitted the potential with the updated training set, which contained 415 configurations. This second potential was then used to conduct additional MD simulations at temperatures 600 K, 800 K, 1,400 K, and 1,600 K, again each for 20 ps with a time step of 1 fs.

Using these simulations we collected 8,000 configurations in total, from which the active learning algorithm selected 142 configurations to be added to the training set. After conducting DFT calculations with these 142 configurations, the results were added to training set. At this stage, the training set contains 557 training data. We then trained another MTP10 potential. We did another round of data generation in the temperature range 600 K – 1600 K using the new potential, but we found that in this step, the algorithm did not sample any additional training data and therefore we stopped the training procedure.

When developing a potential for applications in a certain temperature range, it is a good practice to include training data that belong slightly outside the intended range of application. In the case of FLiBe, the experimental melting point is 732 K, and most of the experimentally available thermophysical properties are in the range 732 K – 1400 K. To make sure that our calculated properties at both ends of this temperature range are reliable, we used a training data that are in the range 600 K – 1600 K. The data at 600 K are for the amorphous solid phase of FLiBe and not the crystalline phase.

It is worth mentioning that the choice of sampling the data from NPT or NVT simulations is arbitrary. The important thing to consider is that the data has sufficient diversity in terms of the supercell size and temperature. Another point to consider is that in the active learning procedure we used a mixture of data obtained from AIMD and from single-point energy calculations by DFT. Specifically, the single-point energy calculations are performed on the atomic configurations that were extracted from MTP-MD simulations. Since the energy and forces depend only on interatomic distances, these properties will not depend on whether the configurations were generated from AIMD or from single-point energy minimization.

Figure 4 shows the parity plots of the energies and forces for the training and Test3 testing set, as predicted by the final potential. The Test3 testing set used in Figure 4 consists of both Test1 and Test2 of Section 3.1 which are uncorrelated to the training data generated through active learning in this section. Table 2 compares the predicted errors of the potential trained by 1400 samples in section 3.1 (MTP10$_{1400}$) and our final potential in this section (MTP10$_{557}$). The predicted errors of both potentials are comparable. This shows that the active learning procedure is effective both in terms of reducing the number of DFT data required to be generated for training and in terms of the training time of MTP, where both were reduced by more than 50%. For the remainder of this work, we use MTP10$_{557}$ to calculate the thermophysical properties of FLiBe.

*Table 2*. *Comparison of RMSEs of energies and forces between MTP10$_{1400}$ and MTP10$_{557}$. The values are in the units of meV/atom for energies, and meV/Å for forces.*

|  | MTP10$_{557}$ | MTP10$_{1400}$ |
|---|---|---|
| Energy RMSE |  |  |
| Test2_600 | 3.0 | 2.9 |
| Test2_1600 | 2.2 | 2.5 |
| Force RMSE |  |  |
| Test2_600 | 48 | 53 |
| Test2_1600 | 62 | 69 |

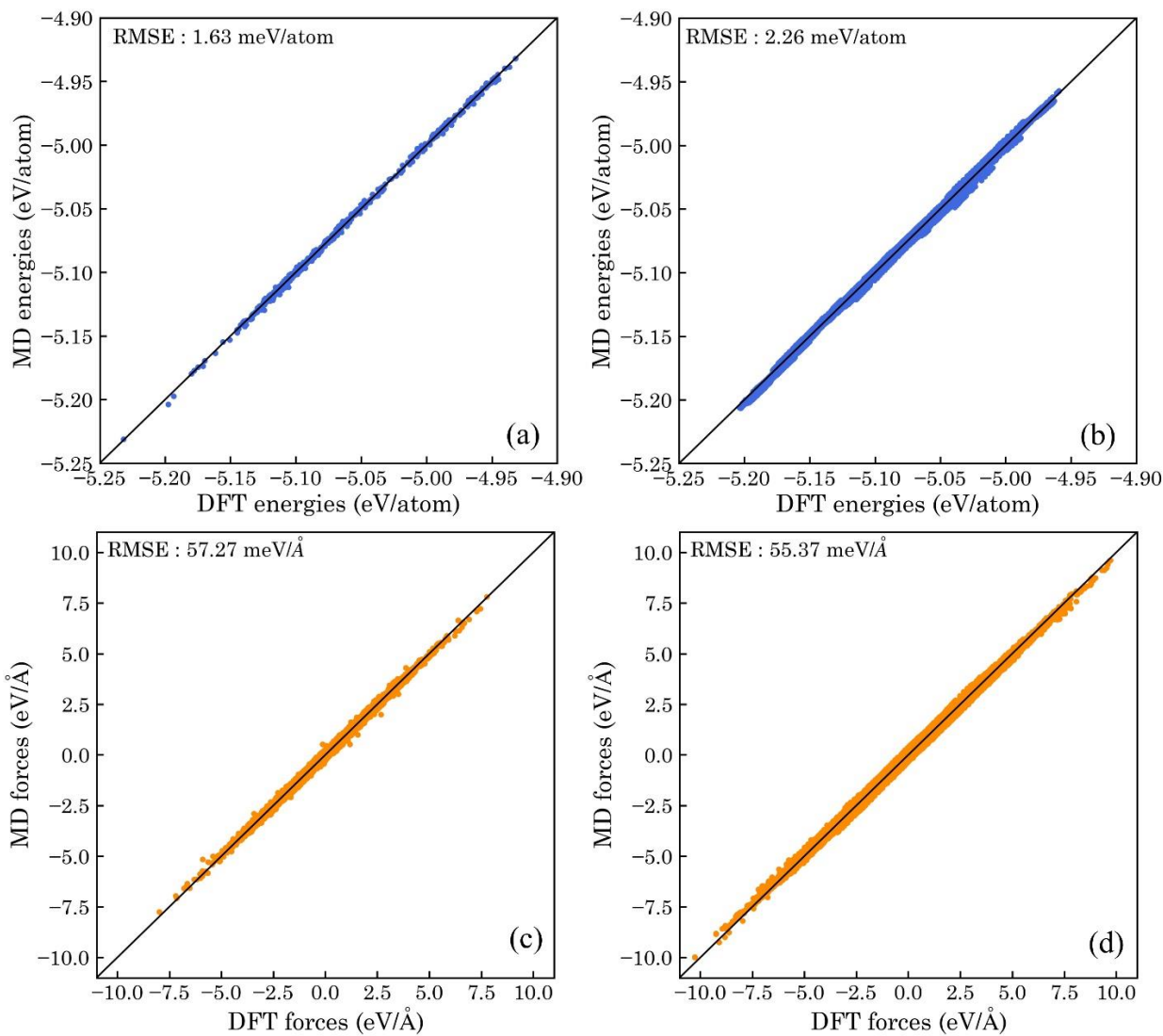

*Figure 4. Parity plots of the energies of the (a) training and (b) Test3 sets and forces of the (c) training and (d) Test3 sets for the MTP10 trained by active learning (MTP10$_{557}$). The diagonal solid line in each figure shows the perfect fit. The Test3 set is the combination of Test1 and Test2 sets consisting of 22600 configurations.*

It is worth mentioning that the overall process of the potential development in this section by active learning, including the initial AIMD simulation for 1 ps, active sampling from the MD trajectories, 457 single point energy calculations using DFT, and 3 sets of MTP potential training were performed in less than 48 hours using a single node with 40 cores based on Intel Xeon Gold 5218R processors.

**3.4. Radial distribution function**

To assess how well the potential developed in this study can predict local structural features, we compared the RDF obtained from MTP MD and our DFT AIMD for $Be^{2+}$-$F^-$, $Li^+$-$F^-$, and $F^-$-$F^-$ at 973 K (see Figure 5). The curves for MTP and DFT almost overlap each other in the range between 0 and 6 Å. RDF for $Be^{2+}$-$F^-$ shows a sharp peak at 1.54 Å and then decays to zero. This suggests a strong bonding between $Be^{2+}$ and $F^-$ within FLiBe. On the other hand, the RDF for $Li^+$-$F^-$ has a wider first peak that does not decay to zero. This result suggests that the first nearest neighbor shell for $Li^+$ and $F^-$ is more diffuse and not as well defined as for $Be^{2+}$ and $F^-$, which further indicated that the bonding between $Li^+$ and $F^-$ in FLiBe is not as strong as between $Be^{2+}$ and $F^-$. Table 3 compares the position of the first RDF peak (which corresponds to the average bond length) and the coordination number of the aforementioned ionic pairs, as obtained from MTP, DFT, IPMD [16], and from experiment [61]. The values of the first RDF peak obtained from MTP are in good agreement with the results obtained from DFT, IPMD, and the experiment. The coordination numbers of $Li^+$-$F^-$ and $F^-$-$F^-$ obtained from MTP and DFT however, do not seem to agree with the experiment. This disagreement likely results from the fact that the location of the first minimum after the peak cannot be confidently identified, especially for $Li^+$-$F^-$ and $F^-$-$F^-$.

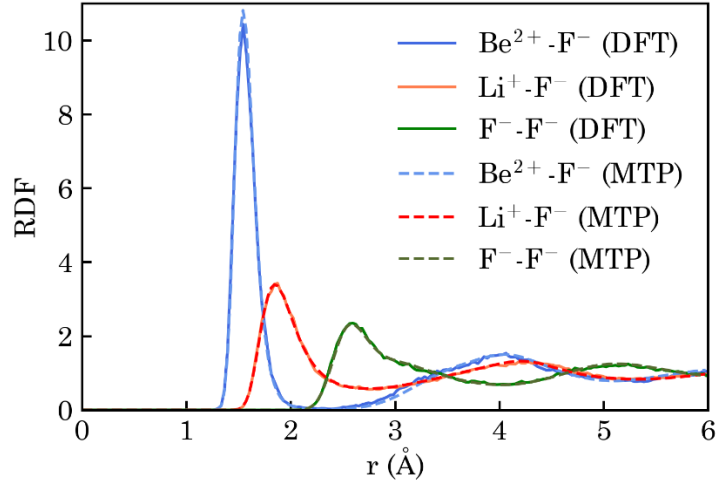

*Figure 5.* *Comparison of the radial distribution functions (RDF) of FLiBe obtained from MTP and DFT at 973 K.*

*Table 3.* *Position of the first peak in RDF shown in Figure 5 and the coordination number. MTP and DFT values are determined in this work, IPMD is from MD simulation with PIM potential [16] and experimental results are from Ref. [61].*

|  | Position of first RDF peak (Å) | | | | Coordination number | | | |
|---|---|---|---|---|---|---|---|---|
|  | MTP | DFT | IPMD | Experiment | MTP | DFT | IPMD | Experiment |
| $Be^{2+}$-$F^-$ | 1.54 | 1.54 | 1.58 | 1.58 | 3.92 | 3.98 | 4.0 | 4.0(3) |
| $Li^+$-$F^-$ | 1.84 | 1.87 | 1.81 | 1.85 | 4.84 | 4.45 | 4.0 | 4 (1) |
| $F^-$-$F^-$ | 2.59 | 2.59 | 2.61 | 2.56 | 11.98 | 11.96 | 11.3 | 8(2) |

In the remainder of the paper, we provide the results of the calculations of the thermophysical properties of FLiBe. For the calculations of density, diffusivity, thermal expansion coefficient, and specific heat capacity we conducted three separate sets of simulations. For the bulk modulus, viscosity, and thermal conductivity due to the computational cost of the simulations, we conducted two sets of simulations, each starting from an uncorrelated supercell. Since the errors in the calculated results are smaller than the size of the data points shown in the figures, we provided the data with their errors in separate tables that can be found in the supplementary data. It should be noted that due to the high computational cost of conducting the MD simulations multiple times the reported sampling errors are based on only a few data points. Given this limited set of independent runs, the sampling error estimates cannot be considered converged and merely serve as a guide on the qualitative scale of the sampling errors and to show that the results are reproducible. Please see the supplementary data for more discussion of the uncertainties.

**3.5. Density**

In Figure 6 we compare the temperature-dependent density calculated from MD simulations with MTP to several results obtained experimentally [45,62–64] and theoretically [10,15,28]. The densities calculated in the current study fall within the range of experimental results. The densities obtained using the van der Waals density functional (vdw-DF) by Nam *et al.* [10] overlap with our results between 800 K and 1,150 K. The densities calculated using IPMD [15] are close to the lower bound of the experimental results. DPMD results are taken from the work of Rodriguez *et al.* [28], where the potential was fitted to DFT data without considering dispersion corrections. The density predicted by the authors is about 10% lower than the lower bound of the experimental results.

We fitted a linear function to our results and obtained the following expression (shown as a black dashed line in Figure 6):

$$\rho = 2.4422 - 4.70 \times 10^{-4} T \ \frac{g}{cm^3}, \quad 600\ K \leq T \leq 1600\ K \qquad Eq.\ 8$$

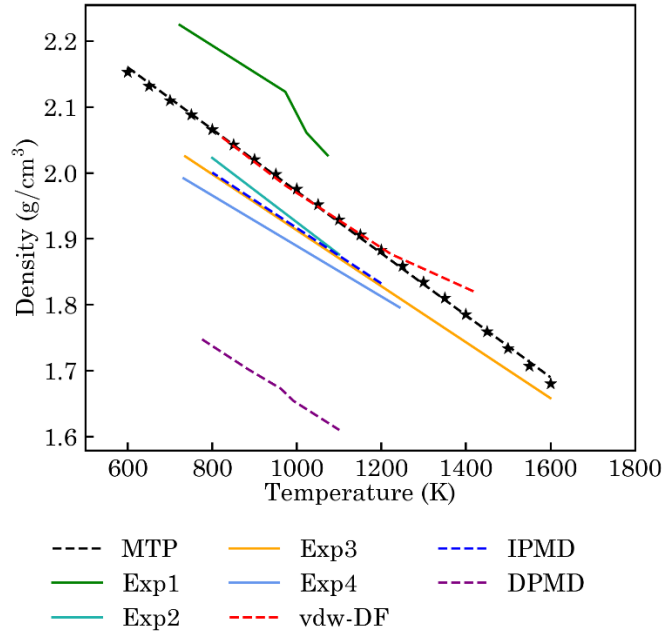

*Figure 6. Temperature-dependent density of FLiBe predicted by MTP compared to values obtained in other experimental (Exp1-Ignat'ev et. al [62], Exp2-Janz [45], Exp3-Humrickhouse and Merrill [63], Exp4-Chen et. al [64]) and theoretical (vdw-DF-Nam et. al [10], IPMD-Smith et. al [15], DPMD-Rodriguez et. al [28]) studies. The experimental results and theoretical predictions are shown with solid and dashed lines, respectively.*

### 3.6. Diffusion

Figure 7 shows the plots of the self-diffusion coefficients of Li, Be, and F versus temperature on a logarithmic scale obtained from MTP (this work), IPMD [16], vdw-DF [10], DPMD [28], and experiment [65]. It can be seen that for all three atomic species, the results obtained using MTP are closer to the experimental results [65] than the results obtained using DPMD [28], which could be due to the lack of accounting for dispersion forces in the development of DPMD. Our calculations are generally in agreement with the experimental results, especially for fluorine (see Fig. 7(c)).

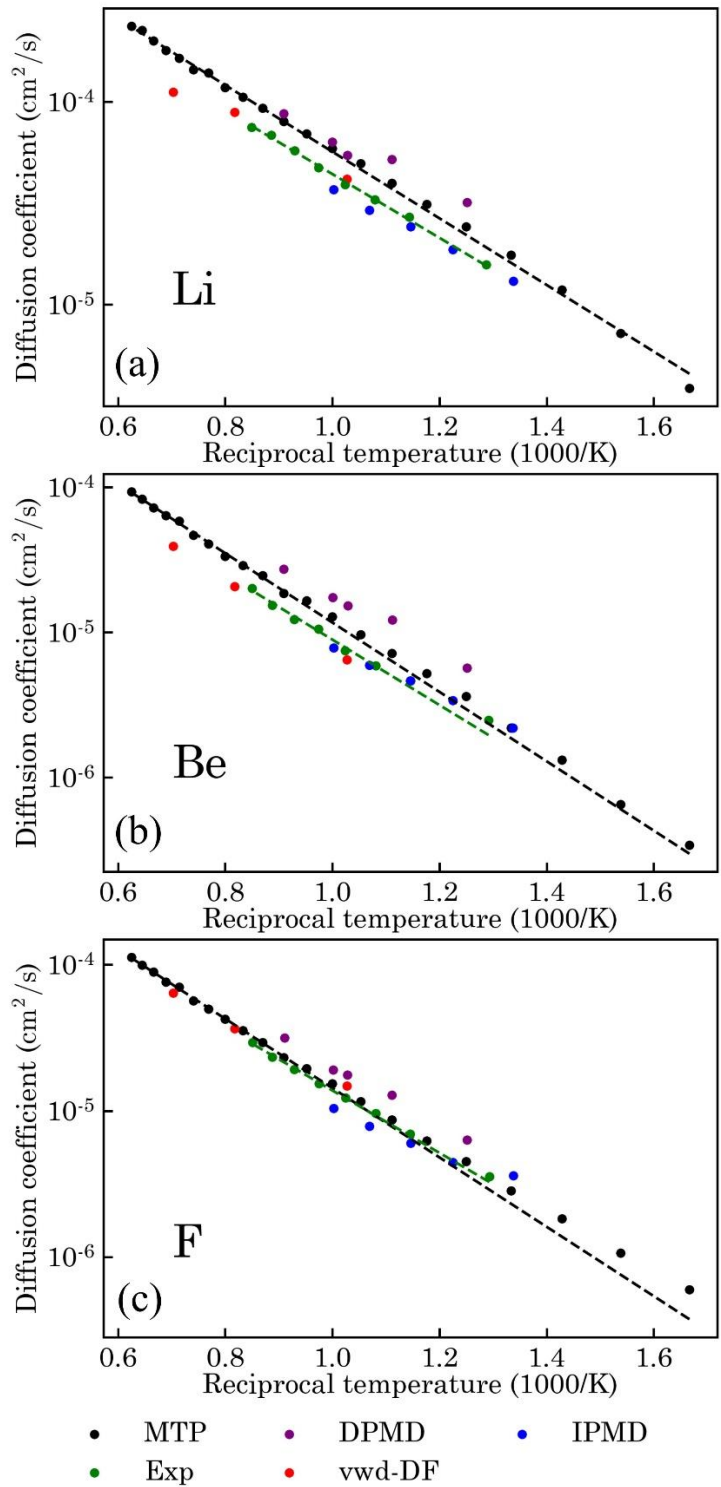

*Figure 7. Temperature-dependent diffusion coefficient of (a) Li, (b) Be, and (c) F calculated from MTP compared to the results using DPMD-Rodriguez et. al* [28]*, IPMD- Salanne et. al*

[16], *vdw-DF-Nam et. al [10] and experiment- Mei et. al [65]. The fitted lines are only shown for MTP and experimental results and are based on Equation 9 with the parameters provided in Table 4.*

Diffusion coefficients were determined by fitting the Arrhenius equation

$$D = D_0 e^{(-\frac{E_A}{RT})} \qquad Eq. 9$$

to data obtained either from MTP or from experiment [65] and the results are shown in Table 4. Due to the few data points reported in the theoretical studies (IPMD [16], DPMD [28], and vdw-DF[10]), the fit would be inaccurate, and therefore we did not fit the Arrhenius equation to those data. In the above equation, $D$ is the self (tracer) diffusion coefficient, $D_0$ is the diffusion prefactor, $E_A$ is the activation energy, $R$ is the universal gas constant, and $T$ is the temperature. The activation energy ($E_A$) corresponds to the slope of the fitted line in Figures 7 (a)-(c) and the diffusion prefactor ($D_0$) corresponds to the intersection of the fitted line with the diffusion axis (vertical axis) when $T \to \infty$ or $\frac{1000}{T} \to 0$.

*Table 4. Parameters of the Arrhenius equation for diffusion. The value of the activation energy ($E_A$) is in the units of kJ/mol, and the prefactor ($D_0$) is in the units of $10^{-6}$ cm$^2$/s. For each method, we provided the temperature range of the reported diffusivity data that were used to fit the parameters.*

|   |       | MTP           | Experiment [65]   |
|---|-------|---------------|-------------------|
|   |       | (600 K– 1600K)| (770 K – 1200 K)  |
| Li| $E_A$ | 31.3          | 30.1              |
|   | $D_0$ | 2459          | 1641              |
| Be| $E_A$ | 45.7          | 43.1              |
|   | $D_0$ | 2889          | 1597              |
| F | $E_A$ | 45.4          | 40.9              |
|   | $D_0$ | 3386          | 1894              |

The activation energies obtained from MTP, and the experiment [65] shown in Table 4 are in good agreement. The prefactors on the other hand show a noticeable difference, which is likely due to the large sensitivity of this value to small differences in the data. This sensitivity can be understood by looking at Figure 7 and considering that a slight deviation between the fitted lines of MTP and the experiment would cause the two fitted lines to intersect the diffusion axis at quite different points when extrapolated to $T \to \infty$ ($\frac{1000}{T} \to 0$)

### 3.7. Thermal expansion coefficient and specific heat capacity

Figure 8 shows the thermal expansion coefficient (TEC) obtained in our work compared to available experimental [45,62–64] and theoretical values [15]. We calculated TECs based on the temperature-dependent density data of each reference, except for the IPMD from Smith *et al.* [15], in which case we used the function the authors provided for TEC. Our results fall within the experimental range. The relationship between TEC and temperature calculated in our work is provided numerically through a simple linear fit below:

$$TEC = 1.8319 \times 10^{-4} + 5.55 \times 10^{-8} \, T \, \frac{1}{K}, \quad 600 \leq T \leq 1600 \qquad Eq.\,11$$

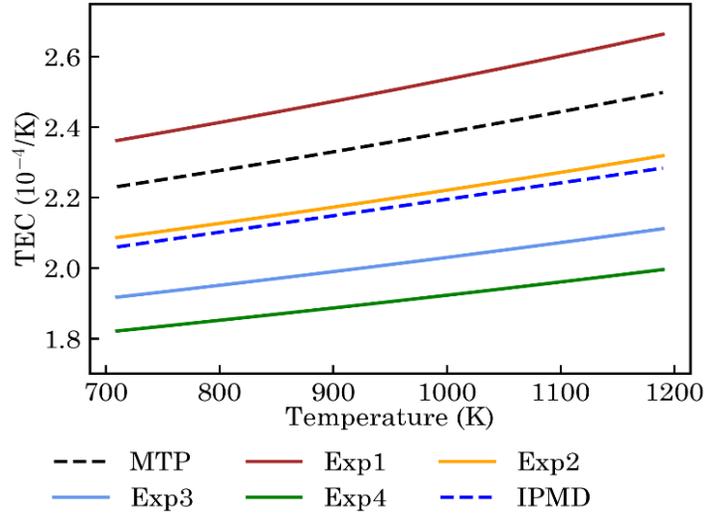

*Figure 8. Temperature-dependent thermal expansion coefficient (TEC) of FLiBe predicted by MTP compared to other experimental (Exp1-Janz [45], Exp2- Humrickhouse and Merrill [63], Exp3-Vaslow and Narten [64], Exp4- Ignat'ev et. al [62]) and theoretical (IPMD-Smith et. al [15]) studies. The experimental results and theoretical predictions are shown with solid and dashed lines, respectively.*

The specific heat capacity ($C_p$) calculated in our work is compared to the experimental works in Table 5. Our result is in good agreement with the experiments.

*Table 5. Specific heat capacity (J/Kg.K)*

| | |
|---|---|
| MTP | 2245 |
| (600 K - 1600 K) | |
| Douglas and Payne [66] | 2347 |
| (745 K - 900 K) | |
| Benes and Konings [67] | 2390 |
| (Temperature not mentioned) | |
| Gierszewski *et al.* [68] | 2380 ± 20% |
| (600 K - 1200 K) | |

### 3.8. Bulk modulus

Figure 9 shows the temperature-dependent bulk modulus calculated in our work, vdw-DF simulations of Nam *et al.* [10], and the experimental work of Cantor *et al.* [69]. The experimental results are based on the compressibility data provided in Ref. [69] with an uncertainty of a factor of 3. Our results fall within the uncertainties of the experiment. Comparing MTP to vdw-DF, the predicted bulk moduli are close at 800 K and begin to deviate as the temperature increases. Considering Figure 6, where the densities of MTP and vdw-DF (calculated at pressure $P=0$) overlap between 800 K and 1150 K, one may expect that the bulk moduli would also follow the same trend. It should be noted that the bulk modulus is a function of the relation between pressure and density. Although the predictions of density at $P=0$ at $T=T_1$ may be the same between MTP (fitted to DFT-D3) and vdw-DF, it is not guaranteed that their predictions would exactly be the same at $P>0$ or $P<0$ at $T=T_1$. In addition, as shown in Figure 6, the density vs. temperature calculated from vdw-DF starts to deviate from its initial linear trend above 1150 K. In contrast, density calculated from MTP (fitted to DFT-D3) shows a linear trend with temperature for the entire range of temperatures considered in this study. It is possible that if the vdw-DF had followed its initial linear trend in density in the entire temperature range, perhaps the bulk modulus calculated from vdw-DF would show a similar trend with temperature as the bulk modulus calculated from MTP.

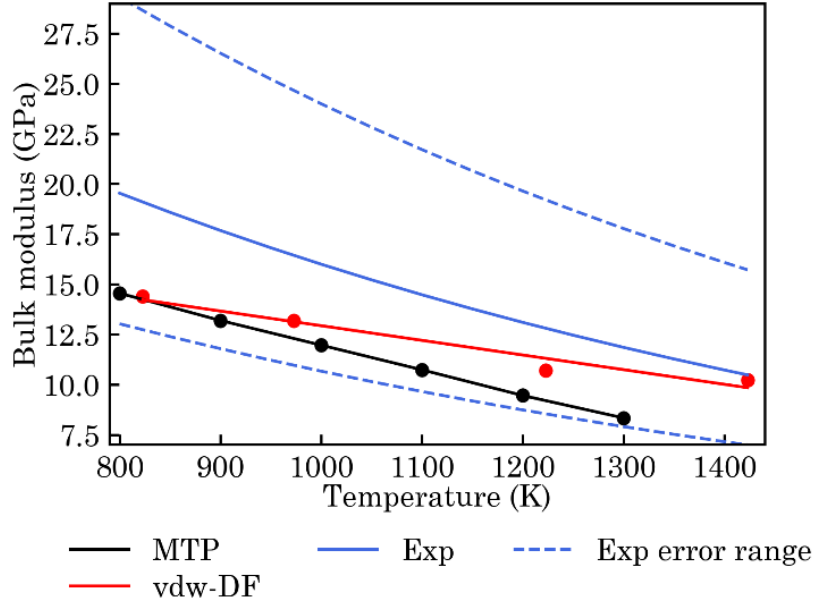

*Figure 9. Temperature-dependent bulk modulus of FLiBe predicted by MTP compared to the experiment-Cantor et. al* [69] *and predictions of vdw-DF-Nam et. al [10].*

### 3.9. Viscosity

Figure 10 shows the temperature-dependent viscosity of FLiBe calculated in this work compared to other experimental [68,70–73] and theoretical [15,28] works. Our results are in excellent agreement with the experimental results of Abe *et al.* [72] and Blanke *et al.* [73]. The predicted results of Smith *et al.* [15] using IPMD are closer to the experimental work of Janz *et al.* [70]. DPMD underestimates the viscosity of FLiBe at lower temperatures. The relationship between viscosity and temperature calculated in our work is as follows

$$\mu = 0.0638 \times Exp\left(\frac{4125}{T}\right) \quad (\text{mPa.s}) \qquad Eq.\,12$$

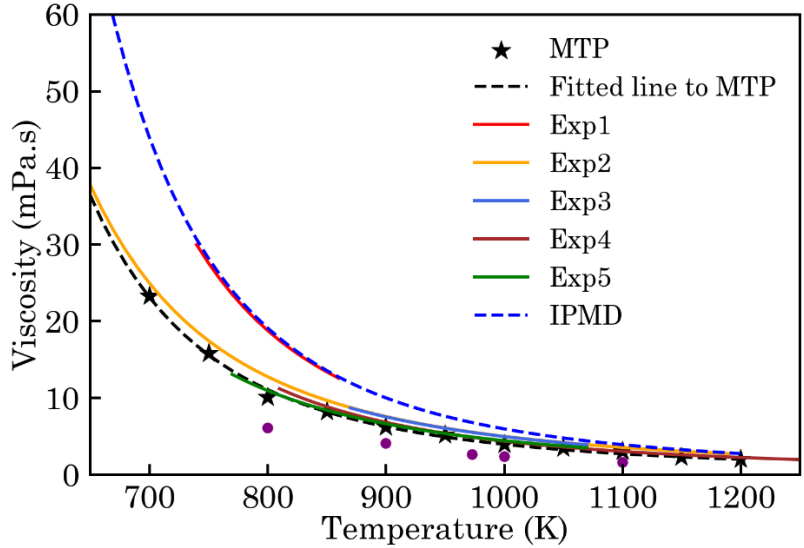

*Figure 10. Temperature-dependent viscosity of FLiBe predicted by MTP compared to other experimental (Exp1-Janz et. al [70], Exp2- Gierszewski et. al [68], Exp3-Cohen and Jones [71], Exp4-Abe et. al [72], Exp5- McDuffie et. al [73]) and theoretical (IPMD-Smith et. al [15], DPMD-Rodrigues et. al[28]) studies. The experimental results and theoretical predictions are shown with solid and dashed lines respectively.*

### 3.10. Thermal conductivity

Figure 11 shows the thermal conductivity of FLiBe calculated in this work and other experimental [68,74] and theoretical [15,28] works. As can be seen, the measured thermal conductivity of each experiment is constant in the temperature range 800 K < T < 1200 K. According to the fitted line to our calculation with MTP, the thermal conductivity increases very slightly from 1.28 to 1.32 in the temperature range 800 K < T < 1200 K and is about 20% higher than the measurement of McDuffie *et al.* [74] (shown as Exp1 in Figure 11). The theoretical work of Smith *et al.* [15] using IPMD also overestimates the thermal conductivity. In their work, the authors argued that in fully dissociated ionic systems, the calculated thermal conductivity deviates from experimental results, especially for lighter alkali cations such as Li. Based on this argument, they obtained the thermal conductivity of LiF using IPMD, calculated the difference between their prediction and experimental results for LiF, and used this difference to correct their calculation of the thermal conductivity of FLiBe. By doing so they lowered their initially predicted thermal conductivity of FLiBe by 16%. The predictions of DPMD underestimate the thermal conductivity at lower temperatures and show an increasing trend with increasing temperature.

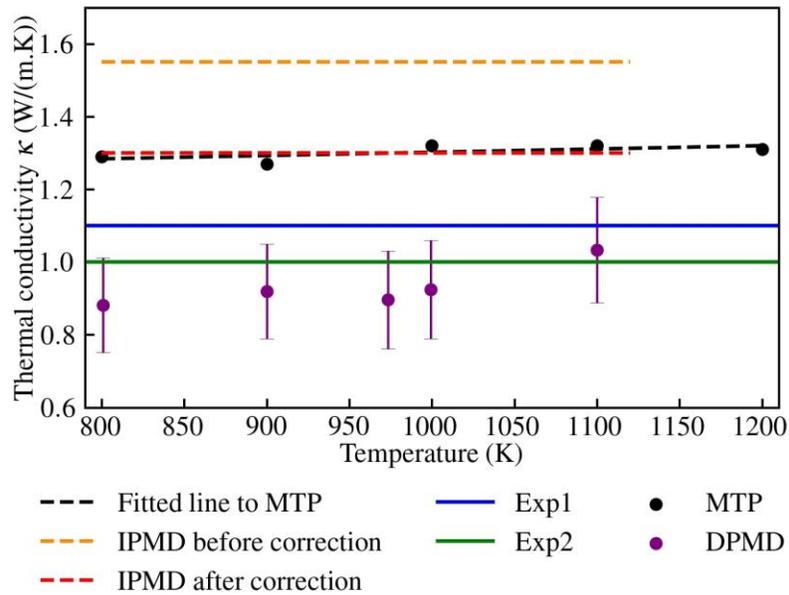

*Figure 11. Temperature-dependent thermal conductivity (κ) of FLiBe predicted by MTP compared to other experimental (Exp1- McDuffie et. al [74], Exp2- Gierszewski et. al [68]) and theoretical (IPMD-Smith et. al [15], DPMD-Rodriguez et. al [28]) studies. The experimental results and theoretical predictions are shown with solid and dashed lines, respectively.*

The difference in the thermal conductivity of molten salts calculated from MTP and determined from experiments is similar to differences reported in other studies on molten salts with MLIPs [53,75,76]. One possible explanation for these differences lies in the inherent approximations of DFT calculations. For example, in a recent work Tisi *et al.*[77] developed a DPMD potential for water trained on DFT-PBE. They calculated the thermal conductivity of water by both AIMD and DPMD simulation and showed that the results are within 5% of each other in the temperature range 400 K – 520 K, but about 60% higher than the experimental values. The authors trained another DPMD based on the more accurate DFT-SCAN and this time the predicted thermal conductivities were around 30% higher than the experimental results. Such approximations inherent to the exchange-correlation functionals carry over into the machine learning based interatomic potentials and affect the properties calculated by MD simulations.

## 4. Conclusion:

In this work, we developed a machine learning potential based on the MTP framework for FLiBe and assessed the performance of the MTP potentials in modeling molten salts. The results showed that training MTP with as low as 600 samples provides accuracies of less than 3 meV/atom for energies and less than 60 meV/Å for forces. The entire process of potential development, including the data generation and the training, can be done in less than 2 days with a single node with 40 cores. The computational cost of MD simulations with MTP in our work is lower than the reported values for equivalent molten salts where other MLIPs such as GAP and NNIP are used. We calculated several thermophysical properties, including radial distribution functions, density, self-diffusion coefficients, thermal expansion, specific heat capacity, bulk modulus, viscosity, and thermal conductivity, of FLiBe and compared the results to the available experimental data. Our predicted properties are generally in very good agreement with experiments and our results suggest that considering van der Waals dispersions during the generation of training data improves the predictions of the developed potentials. Our results demonstrate that MTP potential is viable for modeling the thermophysical properties of molten salts.

**Acknowledgments**

We gratefully acknowledge support from the Department of Energy (DOE) Office of Nuclear Energy's (NE) Nuclear Energy University Programs (NEUP) under award # 21-24582.

*Supplementary data*

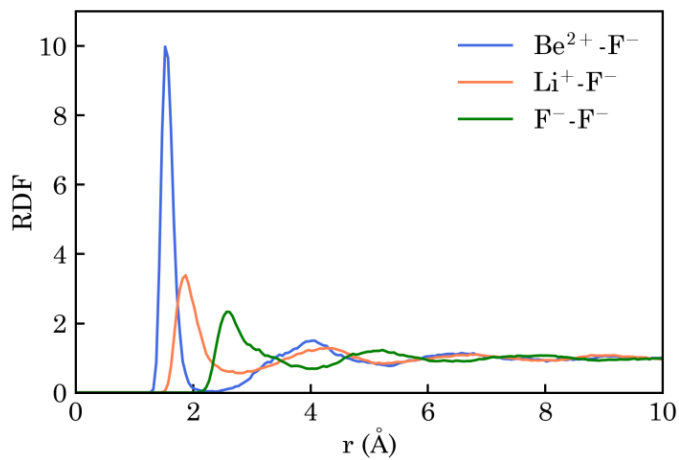

***Figure A.1*** *Radial distribution functions (RDF) of FLiBe obtained from DFT at 973 K. Some structural features are detectable up to 10 Å.*

The reported errors in Tables A.1 to A.10 are sampling errors, which means they are due to our limited sample sizes used to calculate the properties. Our approach of using totally separate independent runs has the advantage of being very straightforward, applicable to any property, and clearly correct in the limit of many independent runs. However, it has the disadvantage of being very slow, as all the data needs to be rerun for each independent case. For this reason, we include just 2-3 independent runs in the tables below. Given this limited set of independent runs the sampling error estimates cannot be considered converged and merely serve as a guide on the qualitative scale of the sampling errors and to show that the results are reproducible. In general, the sampling errors appear to be significantly smaller or at most comparable than both the spread in values between experiments (when available) and the expected errors from DFT, suggesting that our sampling is adequate for the present study. We note that the error for some properties such as density could be analyzed in more details by block averaging over time longer than the correlation length and using the spread in those block averages to report errors. However, the block averaging method cannot be readily applied to much of our data, e.g., bulk moduli, thermal conductivity, viscosity, etc., either because the results are processed through a step of non-linear curve fitting where error propagation is not straightforward or because the property requires all the available data in the simulation to obtain a robust single value. We have therefore decided to report the error consistently using just one method based on separate independent simulations.

A.1. *Calculated density of FLiBe (in gr/cm³) from three separate simulation sets. The density for each simulation is the average of the density of 100 ps. The error is the standard error of the mean* $\left(\sigma_{\bar{x}} = \frac{\sigma}{\sqrt{n}}\right)$ *where σ is the standard deviation in the density of the simulations and n is the number of independent simulations (three).*

| Temperature | Simulation 1 | Simulation 2 | Simulation 3 | Average | Error |
| --- | --- | --- | --- | --- | --- |
| 600 | 2.1528 | 2.1503 | 2.1551 | 2.1527 | 0.0011 |
| 650 | 2.1316 | 2.1329 | 2.1299 | 2.1315 | 0.0007 |
| 700 | 2.1117 | 2.1096 | 2.1083 | 2.1098 | 0.0008 |
| 750 | 2.0883 | 2.0888 | 2.0878 | 2.0883 | 0.0002 |
| 800 | 2.0653 | 2.0665 | 2.0661 | 2.0660 | 0.0003 |
| 850 | 2.0430 | 2.0431 | 2.0427 | 2.0429 | 0.0001 |
| 900 | 2.0205 | 2.0205 | 2.0209 | 2.0206 | 0.0001 |
| 950 | 1.9970 | 1.9983 | 1.9984 | 1.9979 | 0.0004 |
| 1000 | 1.9753 | 1.9751 | 1.9760 | 1.9755 | 0.0002 |
| 1050 | 1.9519 | 1.9519 | 1.9524 | 1.9521 | 0.0001 |
| 1100 | 1.9292 | 1.9294 | 1.9288 | 1.9291 | 0.0001 |
| 1150 | 1.9065 | 1.9069 | 1.9053 | 1.9062 | 0.0004 |
| 1200 | 1.8813 | 1.8831 | 1.8834 | 1.8826 | 0.0005 |
| 1250 | 1.8588 | 1.8594 | 1.8586 | 1.8589 | 0.0002 |
| 1300 | 1.8350 | 1.8341 | 1.8337 | 1.8342 | 0.0003 |
| 1350 | 1.8099 | 1.8097 | 1.8100 | 1.8099 | 0.0001 |
| 1400 | 1.7840 | 1.7863 | 1.7847 | 1.7850 | 0.0006 |
| 1450 | 1.7596 | 1.7583 | 1.7597 | 1.7592 | 0.0004 |
| 1500 | 1.7342 | 1.7344 | 1.7320 | 1.7335 | 0.0006 |
| 1550 | 1.7061 | 1.7070 | 1.7075 | 1.7069 | 0.0003 |
| 1600 | 1.6811 | 1.6801 | 1.6797 | 1.6803 | 0.0003 |

*A.2. Diffusion coefficient of Li in FLiBe (in $10^{-6}$ cm/s) from three separate simulation sets. The error is the standard error of the mean as defined in the caption of Table A.1*

| Temperature | Simulation 1 | Simulation 2 | Simulation 3 | Average | Error |
|---|---|---|---|---|---|
| 600 | 3.93 | 4.01 | 3.64 | 3.86 | 0.09 |
| 650 | 7.09 | 7.16 | 7.30 | 7.18 | 0.05 |
| 700 | 11.83 | 11.38 | 12.21 | 11.80 | 0.20 |
| 750 | 16.62 | 17.98 | 17.82 | 17.47 | 0.35 |
| 800 | 23.56 | 24.14 | 24.65 | 24.12 | 0.26 |
| 850 | 31.09 | 31.50 | 30.93 | 31.17 | 0.14 |
| 900 | 40.48 | 37.67 | 40.80 | 39.65 | 0.81 |
| 950 | 50.82 | 47.45 | 50.19 | 49.49 | 0.84 |
| 1000 | 59.06 | 59.11 | 58.16 | 58.78 | 0.25 |
| 1050 | 65.65 | 69.87 | 72.18 | 69.23 | 1.56 |
| 1100 | 81.50 | 82.24 | 75.85 | 79.86 | 1.65 |
| 1150 | 94.91 | 89.96 | 93.95 | 92.94 | 1.24 |
| 1200 | 104.82 | 104.13 | 105.55 | 104.83 | 0.33 |
| 1250 | 121.22 | 114.05 | 116.29 | 117.19 | 1.73 |
| 1300 | 138.89 | 137.42 | 139.10 | 138.47 | 0.43 |
| 1350 | 146.12 | 138.50 | 145.78 | 143.47 | 2.03 |
| 1400 | 162.90 | 160.45 | 166.11 | 163.15 | 1.34 |
| 1450 | 176.12 | 176.64 | 182.18 | 178.31 | 1.58 |
| 1500 | 206.32 | 194.21 | 195.92 | 198.82 | 3.09 |
| 1550 | 227.16 | 215.53 | 231.18 | 224.62 | 3.83 |
| 1600 | 234.41 | 235.20 | 235.26 | 234.95 | 0.22 |

*A.3. Diffusion coefficient of Be in FLiBe (in $10^{-6}$ cm/s) from three separate simulation sets. The error is the standard error of the mean as defined in the caption of Table A.1*

| Temperature | Simulation 1 | Simulation 2 | Simulation 3 | Average | Error |
|---|---|---|---|---|---|
| 600 | 0.33 | 0.36 | 0.34 | 0.34 | 0.01 |
| 650 | 0.61 | 0.76 | 0.60 | 0.65 | 0.04 |
| 700 | 1.43 | 1.14 | 1.39 | 1.32 | 0.07 |
| 750 | 2.22 | 2.39 | 2.00 | 2.21 | 0.09 |
| 800 | 3.83 | 3.67 | 3.34 | 3.61 | 0.12 |
| 850 | 5.49 | 4.97 | 5.12 | 5.19 | 0.13 |
| 900 | 7.33 | 6.92 | 7.32 | 7.19 | 0.11 |
| 950 | 9.57 | 9.60 | 9.81 | 9.66 | 0.06 |
| 1000 | 13.43 | 12.05 | 13.11 | 12.86 | 0.34 |
| 1050 | 16.90 | 16.23 | 16.40 | 16.51 | 0.16 |
| 1100 | 18.78 | 19.20 | 17.67 | 18.55 | 0.37 |
| 1150 | 24.58 | 24.97 | 24.27 | 24.61 | 0.16 |
| 1200 | 30.18 | 28.35 | 28.46 | 29.00 | 0.48 |
| 1250 | 33.02 | 34.23 | 33.20 | 33.48 | 0.31 |
| 1300 | 45.57 | 36.73 | 39.46 | 40.59 | 2.13 |
| 1350 | 46.18 | 47.46 | 46.13 | 46.59 | 0.36 |
| 1400 | 60.59 | 57.30 | 57.92 | 58.60 | 0.82 |
| 1450 | 65.65 | 62.96 | 63.13 | 63.91 | 0.71 |
| 1500 | 73.02 | 75.93 | 67.03 | 71.99 | 2.14 |
| 1550 | 85.44 | 79.11 | 83.37 | 82.64 | 1.52 |
| 1600 | 88.03 | 93.65 | 97.03 | 92.91 | 2.14 |

*A.4. Diffusion coefficient of F in FLiBe (in $10^{-6}$ cm/s) from three separate simulation sets. The error is the standard error of the mean as defined in the caption of Table A.1*

| Temperature | Simulation 1 | Simulation 2 | Simulation 3 | Average | Error |
|---|---|---|---|---|---|
| 600 | 0.61 | 0.60 | 0.58 | 0.60 | 0.01 |
| 650 | 1.01 | 1.20 | 0.99 | 1.07 | 0.05 |
| 700 | 1.94 | 1.67 | 1.86 | 1.82 | 0.07 |
| 750 | 2.89 | 3.01 | 2.64 | 2.85 | 0.09 |
| 800 | 4.79 | 4.47 | 4.27 | 4.51 | 0.12 |
| 850 | 6.55 | 5.97 | 6.21 | 6.25 | 0.14 |
| 900 | 8.63 | 8.44 | 8.94 | 8.67 | 0.12 |
| 950 | 11.47 | 11.41 | 11.94 | 11.60 | 0.14 |
| 1000 | 15.80 | 14.34 | 15.83 | 15.32 | 0.40 |
| 1050 | 19.64 | 19.15 | 19.56 | 19.45 | 0.12 |
| 1100 | 22.61 | 24.44 | 22.71 | 23.26 | 0.48 |
| 1150 | 29.17 | 29.76 | 29.56 | 29.50 | 0.14 |
| 1200 | 36.59 | 34.67 | 34.99 | 35.42 | 0.48 |
| 1250 | 41.47 | 42.99 | 42.60 | 42.35 | 0.37 |
| 1300 | 51.88 | 47.92 | 49.27 | 49.69 | 0.95 |
| 1350 | 56.87 | 57.66 | 55.24 | 56.59 | 0.58 |
| 1400 | 69.86 | 71.42 | 69.27 | 70.19 | 0.52 |
| 1450 | 76.79 | 73.09 | 78.25 | 76.05 | 1.25 |
| 1500 | 87.29 | 92.92 | 86.73 | 88.98 | 1.61 |
| 1550 | 100.69 | 99.52 | 98.82 | 99.68 | 0.45 |
| 1600 | 115.57 | 109.55 | 112.01 | 112.38 | 1.43 |

*A.5. Parameters of the Arrhenius equation for diffusion from each simulation. The value of the activation energy ($E_A$) is in the units of kJ/mol, and the prefactor ($D_0$) is in the units of $10^{-6}$ cm$^2$/s. The errors are the standard error of the mean as defined in the caption of Table A.1*

|    |       | Simulation 1 | Simulation 2 | Simulation 3 | Average | Error |
|----|-------|--------------|--------------|--------------|---------|-------|
| Li | $E_A$ | 31.38 | 31.13 | 31.49 | 31.33 | 0.09 |
|    | $D_0$ | 2490  | 2374  | 2512  | 2459  | 34.95 |
| Be | $E_A$ | 43.90 | 46.23 | 47.23 | 45.79 | 0.81 |
|    | $D_0$ | 2477  | 2965  | 3224  | 2889  | 178.81 |
| F  | $E_A$ | 45.85 | 45.15 | 45.31 | 45.44 | 0.17 |
|    | $D_0$ | 3532  | 3296  | 3329  | 3386  | 60.24 |

*A.6. The slope and the intercept of thermal expansion coefficient of FLiBe (in $\frac{1}{K}$ and $\frac{1}{K^2}$ respectively) from three separate simulation sets. The error is the standard error of the mean as defined in the caption of Table A.1*

|           | Simulation 1 | Simulation 2 | Simulation 3 | Average | Error |
|-----------|--------------|--------------|--------------|---------|-------|
| Slope     | $5.55 \times 10^{-8}$ | $5.55 \times 10^{-8}$ | $5.56 \times 10^{-8}$ | $5.55 \times 10^{-8}$ | $3.56 \times 10^{-11}$ |
| Intercept | $1.8318 \times 10^{-4}$ | $1.8311 \times 10^{-4}$ | $1.8329 \times 10^{-4}$ | $1.8319 \times 10^{-4}$ | $4.19 \times 10^{-8}$ |

*A.7. Specific heat capacity (in J/Kg.K) from three separate simulation sets. The value for each simulation set is the average of the specific heat at 21 temperatures between 600 K – 1600 K. The error is the standard error of the mean as defined in the caption of Table A.1*

| Simulation 1 | Simulation 2 | Simulation 3 | Average | Error |
|--------------|--------------|--------------|---------|-------|
| 2245 | 2241 | 2249 | 2245 | 1.09 |

*A.8. Bulk modulus of FLiBe (in GPa) from two separate simulation sets. The error is the standard error of the mean as defined in the caption of Table A.1*

| Temperature | Simulation 1 | Simulation 2 | Average | Error |
|---|---|---|---|---|
| 800 | 14.55 | 14.48 | 14.52 | 0.03 |
| 900 | 13.19 | 13.18 | 13.19 | 0.00 |
| 1000 | 11.96 | 11.99 | 11.98 | 0.01 |
| 1100 | 10.73 | 10.62 | 10.68 | 0.04 |
| 1200 | 9.45 | 9.49 | 9.47 | 0.01 |
| 1300 | 8.33 | 8.29 | 8.31 | 0.01 |

*A.9. Viscosity of FLiBe (in mPa.s) from two separate simulation sets. The error is the standard error of the mean as defined in the caption of Table A.1*

| Temperature | Simulation 1 | Simulation 2 | Average | Error |
|---|---|---|---|---|
| 700 | 21.88 | 24.70 | 23.29 | 1.00 |
| 750 | 16.74 | 14.85 | 15.80 | 0.67 |
| 800 | 10.70 | 9.48 | 10.09 | 0.43 |
| 850 | 8.44 | 8.06 | 8.25 | 0.13 |
| 900 | 6.16 | 6.37 | 6.27 | 0.08 |
| 950 | 5.35 | 5.08 | 5.21 | 0.10 |
| 1000 | 3.99 | 3.90 | 3.95 | 0.03 |
| 1050 | 3.33 | 3.59 | 3.46 | 0.09 |
| 1100 | 2.90 | 3.16 | 3.03 | 0.09 |
| 1150 | 2.29 | 2.36 | 2.32 | 0.03 |
| 1200 | 2.11 | 2.01 | 2.06 | 0.04 |

*A.10. Thermal conductivity of FLiBe (in W/m.K) from two separate simulation sets. The error is the standard error of the mean as defined in the caption of Table A.1*

| Temperature | Simulation 1 | Simulation 2 | Average | Error |
|---|---|---|---|---|
| 800 | 1.30 | 1.27 | 1.29 | 0.01 |
| 900 | 1.27 | 1.27 | 1.27 | 0.00 |
| 1000 | 1.36 | 1.27 | 1.32 | 0.03 |
| 1100 | 1.23 | 1.41 | 1.32 | 0.06 |
| 1200 | 1.29 | 1.32 | 1.31 | 0.01 |